\RequirePackage{fix-cm}
\RequirePackage{amsmath}
\documentclass[smallextended]{svjour3}
\smartqed  
\usepackage{graphicx}
\usepackage{bbm}
\usepackage{xspace}
\usepackage{amsfonts}
\usepackage{siunitx}
\usepackage{color}
\begin{document}

\title{Peri-Net-Pro: The neural processes with quantified uncertainty for crack patterns}

\author{Moonseop Kim \and Guang Lin}

\institute{*Corresponding author \at 
Department of Mathematics, Department of Mechanical Engineering, Department of Statistics(Courtesy), Department of Earth, Atmospheric, and Planetary Sciences(Courtesy), Purdue University, West Lafayette, IN 47906-2045, \\ 
Tel.: +1 765 494 1965\\
\email{guanglin@purdue.edu (Guang Lin*) \and kim2122@purdue.edu (Moonseop Kim))         
}
}

\date{Received: date / Accepted: date}
\maketitle

\begin{abstract}

This paper uses the peridynamic theory, which is well-suited to crack studies, to predict the crack patterns in a moving disk and classify them according to the modes and finally perform regression analysis. In that way, the crack patterns are obtained according to each mode by Molecular Dynamic (MD) simulation using the peridynamics. Image classification and regression studies are conducted through Convolutional Neural Networks (CNNs) and the neural processes. First, we increased the amount and quality of the data using peridynamics, which can theoretically compensate for the problems of the finite element method (FEM) in generating crack pattern images.
Second, we did the case study for the PMB, LPS, and VES models that were obtained using the peridynamic theory. Case studies were performed to classify the images using CNNs and determine the PMB, LBS, and VES models' suitability. Finally, we performed the regression analysis for the images of the crack patterns with neural processes to predict the crack patterns.
In the regression problem, by representing the results of the variance according to the epochs, it can be confirmed that the result of the variance is decreased by increasing the epoch numbers through the neural processes. 
Therefore, the result of the training gradually improves, and the ranges of the variance are reduced to less than 0.035.
The most critical point of this study is that the neural processes make an accurate prediction even if there are missing or insufficient training data.
The results show that if the context points are set to $10$, $100$, $300$, and $784$, the training information is deliberately omitted such as context points of 10, 100, and 300, and the predictions are different when context points are significantly lower. However, when comparing the results of context points 100 and 784, the predicted results appear to be very similar to each other because of the Gaussian processes in the neural processes. Therefore, if the training data is trained through the neural processes, the missing information of training data can be supplemented to predict the results.

\end{abstract}

\section{Introduction}
\label{sec:1}
A significant problem to solve in solid mechanics is associated primarily with the construction of discontinuities in areas that do not exist. A typical example of the formation of discontinuities is a mechanical component failure due to cracking and breakage caused by a dynamic load, high-temperature gradient, shock, explosion, etc. Occasionally, due to human negligence and mistakes in daily life, people's possessions are damaged causing a breakdown. In addition, automobile accidents can cause cracks in the car body. Likewise, in the DOE (Department of Energy), the blades of the thermoelectric power plant and wind power plant can form cracks due to the influence of high temperature or wind, and nuclear reactors are damaged by nuclear fission. Research on the cause of damage to objects with motility is actively underway. Many studies on fracture utilize the Finite Element Method (FEM).
However, FEM has a fatal error. First, the FEM is calculated based on partial differential equations, but there are no partial derivatives on the crack surface. Also, only the approximate solution can be obtained in the calculation of the FEM; the result depends on the mesh, and the user errors can destroy the results. 

To compensate for these defects, we used peridynamics. peridynamic theory was first introduced by S. Silling \cite{Ref1,Ref2,Ref3}. By applying integral equations directly to the surface of the crack, peridynamics can compensate for the absence of partial derivatives and compensate for the fatal error of the FEM. From this point of view, peridynamics is the most appropriate theory for studying cracks or damage. peridynamic theory is applied not only to cracks but also to various materials in which cracks are generated such as membranes and nanofiber \cite{Ref4}, composites, and brittle materials \cite{Ref5} and prediction of viscoelastic materials \cite{Ref6}. The MD simulation tools of LAMMPS \cite{Ref7} and Peridigm \cite{Ref8} are based on peridynamic theory and have been used extensively in crack research, especially in the application of various material properties to crack simulations. For instance, PMB \cite{Ref9}, LPS \cite{Ref10}, and VES \cite{Ref11} models are set up in the MD simulation tool.
However, the MD simulation produces remarkable increasing computational times as its size increases. 

One way to diminish computational cost is to apply deep learning methods. In recent years, Convolutional Neural Networks (CNNs) has been shown to provide better performance and accuracy than classic fully connected networks, particularly for highly structured data applications such as peridynamic models \cite{Ref12}. These networks have the advantage of being able to pre-train, learn the data sets generated from highly accurate models based on peridynamic theory, and encode the process of the model into a neural network. This approximate model can be used to produce immediate results through the training and testing process. Our novel contribution in the present study is three-fold: 
\begin{enumerate}
\item Neural processes (NPs) \cite{Ref13} are employed to predict and quantiy the uncertainties for crack patterns. NPs combine both the advantages of a neural network and the advantages of a Gaussian process with the data obtained from peridynamic theory. We demonstrate through numerical results that NPs can greatly improve the efficiency of computation during the training and evaluation resulting in faster and more accurate results. 

\item We demonstrate that NPs can quantify the uncertainties for predicting crack patterns. Through the training process, the variance was greatly reduced as increasing the epoch number, which indicate the convergence in the hyperparameter estimation of NPs. It can serve as a good indicator on when the training process is finished.    
\item Finally, in this study, CNN and neural processes are applied based on the data obtained from the peridynamic theory. We are able to enhance the peridynamic theory even more precisely through the case study for the characteristic of materials. By applying NPs, we can speed up the calculation speed using the trained data and finally reduce the computational cost.
\end{enumerate}
This paper is constituted as follows. In section 2, we will briefly discuss peridynamic models (PMB, LPS, VES). This study used the neural processes incorporating a gaussian process into a neural network in order to be predictable despite the loss of input data information. Section 3 will introduce the neural processes. In Section 4, we will explain how to obtain input data for training using three peridynamic models. Finally, in section 5, we will determine how well input data obtained from three peridynamic models are predicted through CNNs, and we will use the neural processes to infer the value of variance as the epoch increases and examine how accurate it is predicted through the neural processes.

\section{The peridynamic model (PMB, LPS, VES)}
\label{sec:2}

\begin{figure}
\includegraphics[scale=0.8]{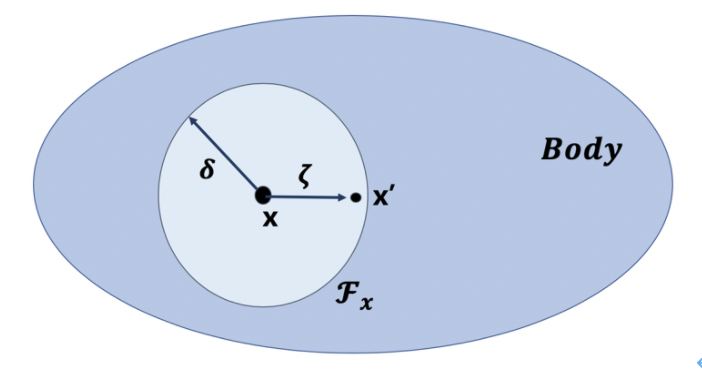}
\caption{The composition of peridynamic model.}
\label{fig:1}       
\end{figure}

In peridynamics, all points in the material are connected by bonds. Fig. 1 shows the composition of the peridynamic model. Suppose that there is an x in the whole body. When we denote the circle with radius is $\delta$ (horizon) centered on x, all the points in it are connected, and this is called the family of x. Therefore, peridynamics is the continuation of the classical continuum mechanics. The most crucial point of peridynamics is that it uses an integral equation. In contrast, FEM uses the partial differential equation for crack analysis. However, there are no partial derivatives on the crack surface. Therefore, peridynamics can solve the absence of partial derivatives by applying integral equations directly to the surface of the crack. From this point of view, peridynamics is the most appropriate theory for studying cracks or damage. The following is the equation of motion based on the peridynamic model. The results are as follows.

\begin{equation}
\rho(x) \, \ddot{u}(x,t) \, \, = \, \, \int_{\mathcal{F}_x} T\left(u(x',t)-u(x,t),x'-x\right) \, \, dV_{x'} \, \, + \, \, b(x,t), \quad t\geq0
\end{equation}

\noindent The left term represents the equation of motion, $\rho$ represents the mass density, and $\ddot{u}$ indicates the acceleration. In the right term, $\mathcal{F}_x$ is the family of x, T is the pairwise force function of the bond which connects point x and $x'$. u is the displacement of the point x according to the time domain. Finally, b represents the external body force density. The following definitions of $\zeta$ and $\eta$ are needed to define the conservation of linear momentum and conservation of angular momentum. $\zeta$ is the position vector in the initial arrangement and $\eta$ is the relative displacement vector.

\begin{equation}
\zeta = x'- x
\end{equation}

\begin{equation}
\eta = u'- u
\end{equation}

\noindent $\zeta$ + $\eta$ indicates the current relative position vector between the points. The T function follows the two properties:

\begin{equation}
T(-\eta,-\zeta)= -T(\eta, \zeta) \quad\forall\eta,\zeta
\end{equation}

\begin{equation}
\quad(\zeta + \eta)\times T(\eta, \zeta)=0 \quad\forall\eta,\zeta    
\end{equation}

\noindent Eq. (4) and (5) represent the conservation of linear momentum and the conservation of angular momentum respectively. Conditions of T satisfy the conservation of linear momentum and the conservation of angular momentum. 
In this study, the crack patterns were formed by applying the above-mentioned peridynamic theory. PMB, LPS, and VES models were applied to do the case study for crack patterns according to the material types. First, the PMB model is a Prototype Micro-elastic Brittle model that is evolved from a model of isotropic and micro-elastic. The strength of the bond depends only on bond stretch s. The equation is as follows:

\begin{equation}
y=|\zeta+\eta|
\end{equation}

\begin{equation}
s=\frac{|\zeta+\eta|-|\zeta|}{|\zeta|}=\frac{y-|\zeta|}{|\zeta|}
\end{equation}

\noindent s is the bond stretch and consists of the current relative position vector between the points and the position vector in the initial arrangement. The definition of PMB model is as follows.
\begin{equation}
F(y(t),\zeta)=g(s(t,\zeta))\mu(t,\zeta)
\end{equation}
\begin{equation}
g(s)=cs \quad\forall s
\end{equation}
\begin{equation}
  \mu(t,\zeta) = \left \{
  \begin{aligned}
    &1, && \textrm{if} \quad s(t',\zeta)<s_0 \quad \textrm{for all} \quad 0 \leq t' \leq t\\
    &0, && otherwise
  \end{aligned} \right.
\end{equation}

\noindent The function $F$ is a product of g and $\mu$ functions, and the g function is a linear scalar-valued function consisting of c and s, which are spring constants and bond stretch, respectively. The $\mu$ function, a component of the F function, as a function of time and position vector, and has 0 and 1 depending on the conditions in Eq. (10). If the bond stretch is less than a critical bond stretch $s_0$, u is 1, and otherwise, it is defined as 0. c, and $s_0$ are as follows.

\begin{equation}
c = \frac{18K}{\pi\delta^4}
\end{equation}

\begin{equation}
s_0=\sqrt{\frac{10G_0}{\pi c\delta^5}}=\sqrt{\frac{5G_0}{9 K\delta}}
\end{equation}

\begin{equation}
G_0=\frac{\pi cs_0^2\delta^5}{10} 
\end{equation}

 \noindent The spring constant c consists of bulk modulus K and $\delta$ (horizon boundary in peridynamic model), while critical bond stretch $s_0$ is a function of shear modulus $G_0$ and horizon $\delta$. Finally, shear modulus $G_0$ consists of spring constant c, and critical bond stretch $s_0$ and horizon $\delta$ in the boundary. The scalar force state of the PMB model is as follows:
 
\begin{equation}
\underbar{t}_{PMB} \, = \, \frac{1}{2}\frac{18K}{\pi\delta^4} \, {\frac{||\eta+\zeta||-||\zeta||}{||\zeta||}}
\end{equation}

\noindent Finally, briefly describing the LPS and VES models, the elastic properties of both models are defined by the bulk modulus and the shear modulus in conjunction with the horizon boundary. However, the VES model requires additional relaxation parameters and time constants. Based on the above description, the scalar force state of the LPS model and the VES model can be described as follows. For the LPS model,

\begin{equation}
\underbar{t}_{LPS} \, = \, -\frac{3K\theta}{m} \, \underbar{$\omega$}\underbar{x} + \alpha\underbar{$\omega$}\underbar{$e^d$}
\end{equation}

\noindent where m, $\theta$, e and $e^d$ are weighted volume, dilatation, extension state, and deviatoric extension state respectively. The bulk modulus is K and the shear modulus G related term $\alpha$ = $\frac{15G}{m}$ \cite{Ref2}. 

\begin{equation}
\underbar{t}_{VES} \, = \, -\frac{3K\theta}{m} \, \underbar{$\omega$}\underbar{x} + (\alpha_\infty+\alpha_i)\underbar{$e^d$} - \alpha_i\underbar{$\omega$}\underbar{e}^{db(i)}
\end{equation}

\noindent For the viscoelasticity model, the key constituent is the decomposition of the scalar extension state into dilatation and deviatoric parts, as well as the additive decomposition of the deviatoric extension state into elastic $e^d$ and back extension $e^{db(i)}$ parts. $\alpha$ = $\alpha_{\infty}$+$\alpha_i$ and 0 $<$ $a_i$ $<$ $\frac{15\mu}{m}$ \cite{Ref11,Ref14}.

\section{The Neural Processes}
\label{sec:3}

This section introduces the neural processes. Neural networks are nonlinear functions that are very easy to train. 
Gaussian processes \cite{Ref15} provide the stochastic framework for learning the distribution of a wide range of nonlinear functions. Thus, in limited data, Gaussian processes can capture probabilistic nature and uncertainty. However, the neural networks are computationally much more scalable than the Gaussian processes as neural networks can handle the vast amounts of data. Therefore, The model of neural processes combines the advantages of the neural networks and the Gaussian processes and so that they can compensate for each other's disadvantages. Fig. 2 represents the diagram of neural processes. The process of neural processes is examined through a diagram. The massive flow of neural processes is as follows. Information flows from the context points through the potential space z to new predictions with target points. z is a random variable that makes neural processes a probabilistic model and captures uncertainty about the function. Once the model has been trained, the posterior distribution of z can be used before making predictions at test time.
In summary, given data is largely divided into context points and target points. It predicts $y_T^*$ using the given context and target points. To learn more about the procedure of neural processes, first, context points pass through the neural networks h to produce their respective representations $r_1$, $r_2$, ... $r_c$. Next, all the $r$'s from each context points are aggregated to obtain a single r. This r is used to parameterize the distribution of z. 

\begin{figure}
\includegraphics[scale=0.35]{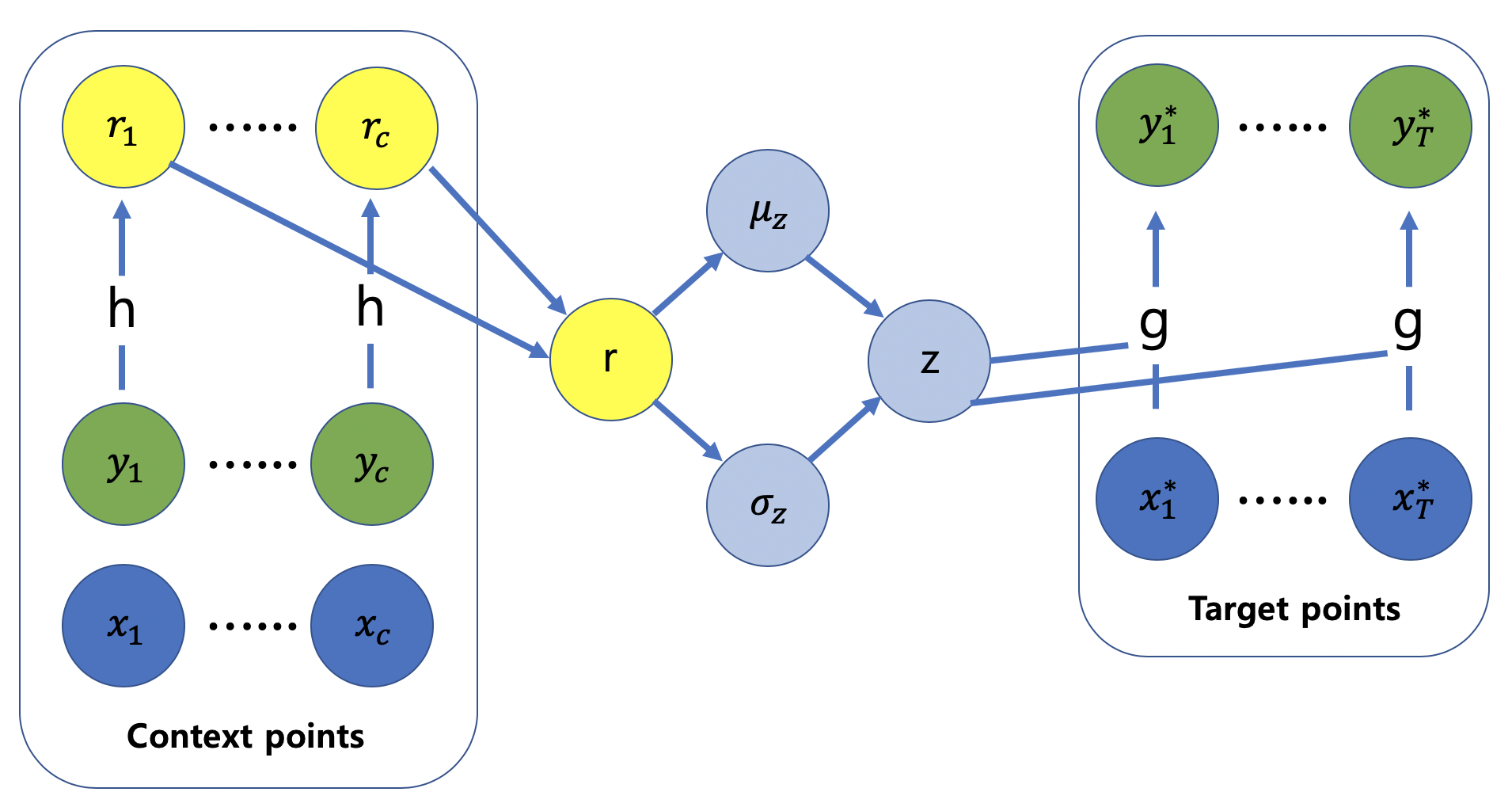}
\caption{The diagram of the neural processes.}
\label{fig:2}       
\end{figure}

\begin{equation}
p(z|x_{1:c},y_{1:c}) = N(\mu_z(r), \sigma_z^2(r))
\end{equation}

\noindent Finally, to obtain the prediction point $y_T^*$, z is sampled, and the result is concatenated with the target point $x_T^*$ to obtain a predictive distribution of $y_T^*$ by passing through a neural network decoder (g). The inference of neural processes is performed in the variational inference framework. Inferring the approximate posterior q(z$|$context, target), we can evaluate the prediction p($y_T^*$ $|$z, $x_T^*$). The approximate posterior q(z$|$context, target) is chosen for prediction $y_T^*$ to use the neural networks h (encoder) and by using the same h, to map the context points and the target points in order to obtain the aggregated r, which in turn is mapped to $\mu_z$ and $\sigma_z$. Lastly, the parameters of the encoder (h) and decoder (g) are trained by the ELBO.

\begin{equation}
ELBO = E_{q(z|c,t)}\footnote{q(z$|$c,t) represents the q(z$|$context,target).}\biggl[\displaystyle\sum_{t=1}^{T} \log p(y_t^*|z,x_t^*) \\
+\log\frac{q(z|context)}{q(z|context,target)}\biggr]
\end{equation}

\noindent The first term in brackets is the expected log-likelihood of the target set. It is evaluated by the first sampling z $\sim$ q(z$|$context, target). The second term of the ELBO has a normalization effect, which is a negative KL divergence between q (z$|$context) and q(z$|$context, target). This term indicates a summary of the contexts to be not too far from the summary of the targets.

\section{Preparation of the data using the peridynamic theory and models}
\label{sec:4}

In this study, peridynamics was applied to obtain the crack patterns of a moving disk and LAMMPS was used for the simulations. 
PMB, VES, and LPS models were used for the case study of the peridynamic model in order to apply deep machine learning. 
The crack patterns were penetrated by the spherical indenter perpendicularly to the disk moving in the x, y, and z directions. 
In this study, data were obtained by changing 4 parameters. 
1) Hitting points were set by increasing and decreasing the x and y coordinates around the center of the disk, and the disk was penetrated by the indenter evenly throughout the cylindrical disk. 
2) The parameters of velocity were changed to 100m/s and 100.1m/s with a slight difference. 
The reason for this was to find out how to classify the velocity of subtle differences when classifying the modes by applying deep machine learning. 
3) The radius of the indenter was adjusted by 0.007m and 0.008m. 4) To obtain the crack pattern of the moving disk, not the static disk, we changed the moving direction of disk x, y, and z. 

\begin{figure}
\includegraphics[height=5.5cm]{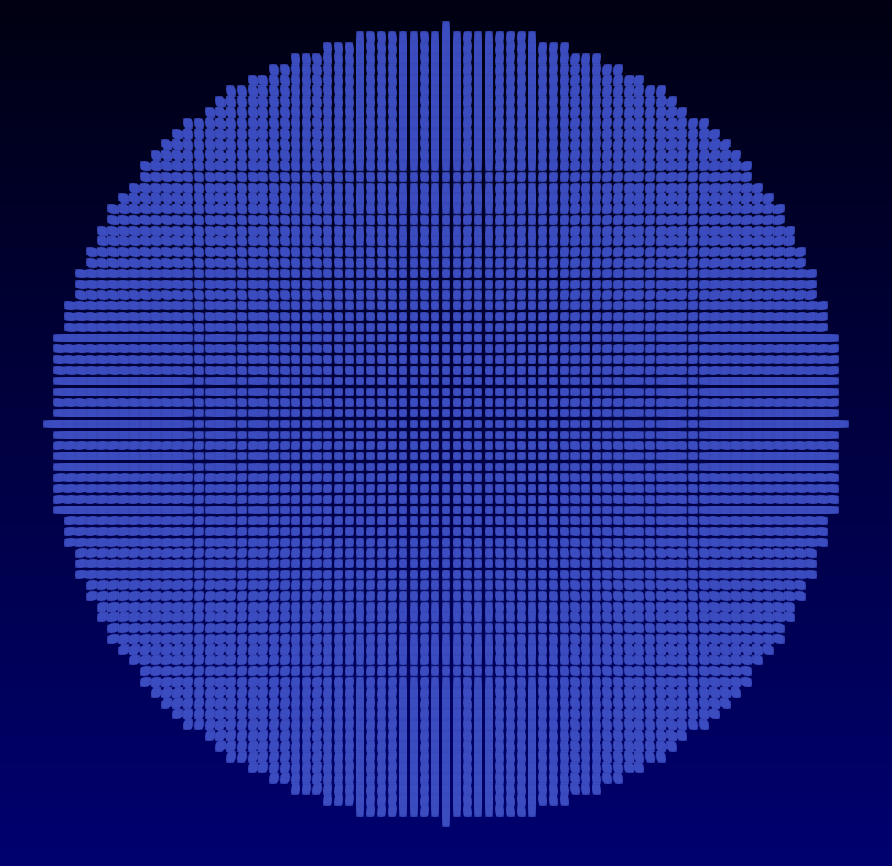}
\label{fig:3.1}
\includegraphics[height=5.5cm]{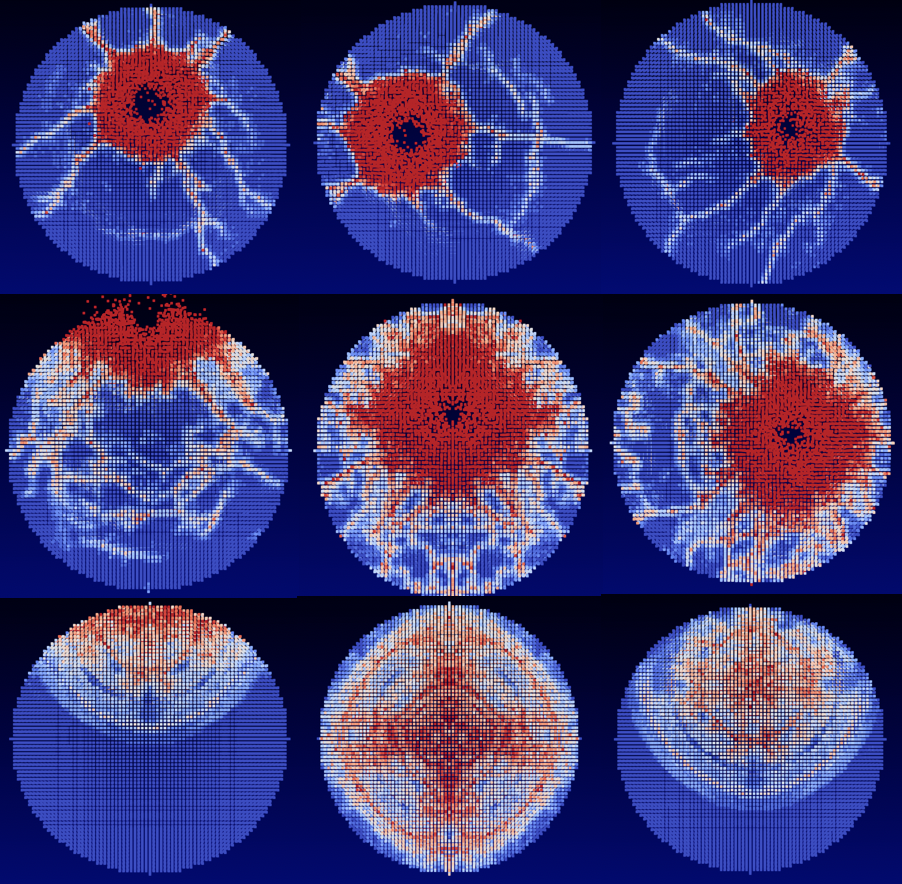}

\caption{Left side : Initial disk, Right side: Crack patterns (LPS, PMB, VES).}
\label{fig:3.2}
\end{figure}

\noindent Finally, the cylindrical disk was composed of 103,110 particles with a radius of 0.037 m and a thickness of 0.0025 m. Each particle has a volume fraction of $V_{i}$ = 1.25 x 10$^{-10}$m$^3$ and the density of the disk material is 
$\rho$ = 2200 kg /m$^3$.
So far, three changed parameters have been applied to the moving disk, and the disk was generated by the PMB model, VES model, and LPS model for the case study. First, to apply the PMB model, `peri / pbs' pair style was used. The pair constants for the simulations have been specified as follows:
c = 1.6863 $\times$ 10$^{22}$, horizon $\delta$ equal to 0.0015001, $s_{00}$ = 0.0005, and $\alpha$ = 0.25. c is the spring constant for peridynamic bonds, the horizon $\delta$ is a cutoff distance for non-local interactions, the constants $s_{00}$ and $\alpha$ correspond to material-dependent parameters for the critical bond stretch.  
Second, we used `peri / lps' pair style for the LPS model which consists of bulk modulus (K), shear modulus (G), horizon, $s_{00}$, and $\alpha$. Bulk modulus (14.9 $\times$ 10$^{9}$), shear modulus (14.9 $\times$ 10$^{9}$), horizon $\delta$ (0.0015001), $s_{00}$ (0.0005) and $\alpha$ (0.25) were used to generate the crack patterns of the LPS model. Finally, to set up the VES model, `peri / ves' pair style consisting of bulk modulus (K), shear modulus (G), horizon, $s_{00}$, $\alpha$, m-lambdai, and m-taubi was used. m-lambdai and m-taubi represent the viscoelastic relaxation parameter and time constant, respectively. 
Lastly, we generated crack patterns for the VES model by applying the bulk modulus = 14.9 $\times$ 10$^{9}$, shear modulus = 14.9 $\times$ 10$^{9}$, horizon = 0.0015001, $s_{00}$ = 0.0005, $\alpha$ = 0.25, m-lambdai = 0.5, and m-taubi = 0.001. Dataset was created for deep machine learning, and the models and simulation settings have been studied for the case studies. We calculated the simulations for 1,000 time-steps in one simulation and stored the output as a dump file, assuming that 1000 time steps were meaningful by tuning the simulation step by step for the case model to obtain the more sophisticated crack patterns. Finally, the pizza.py toolkit \cite{Ref16} was used to change the EnSight data format to visualize the crack patterns using the dump files, then apply it to the Paraview \cite{Ref17,Ref18} tool to visualize the crack patterns. In Fig. 3, the figure on the left is the initial model of the disk before the crack patterns were formed. On the right side, the first row, the middle row, and the last row are the crack patterns that were generated using the PMB, LPS, and VES models respectively.

\section{Method and Results}
\label{sec:5}

This study used supervised machine learning with convolutional neural networks (CNNs) to classify the crack patterns in cases according to PMB, LPS, and VES models and 2-D regression problem with the Neural Processes (NPs). We have applied the deep machine learning method to make MD simulation using peridynamics more efficient, and we also have investigated how accurately the peridynamic model can be predicted through the deep machine learning method. In this section, we study how to define the data set to perform supervised machine learning, how to label the train, validation, test data sets, and how to set up the structures in the CNNs and the NPs. 
Finally, after training, validation, and testing, the loss function is calculated to determine how well the training and validation tests are performed. We will proceed with case studies through the success rate and the result of how accurately we predicted.

\subsection{The preparation of the data for the classification}
\label{sec:5.1}

\begin{table}[ht]
\centering
\caption{12 modes of the input image data for classification.}
\label{tab:1}       
\begin{tabular}{c|c}
\hline
\hline
Mode & 4variables \\
\hline
 & Radius of indenter (r) = 0.007, 0.008m \\
 & Velocity of indenter (v) = 100, 100.1m/s \\
 & Hitting location of disk (x, y)  \\
 & Moving direction of the disk (x, y, z) \\
 \hline
Mode1 & r=0.007m, v=100m/s, Moving direction (x) \\
Mode2 & r=0.007m, v=100.1m/s, Moving direction (x) \\
Mode3 & r=0.007m, v=100m/s, Moving direction (y) \\
Mode4 & r=0.007m, v=100.1m/s, Moving direction (y) \\
Mode5 & r=0.007m, v=100m/s, Moving direction (z) \\
Mode6 & r=0.007m, v=100.1m/s, Moving direction (z) \\
Mode7 & r=0.008m, v=100m/s, Moving direction (x) \\
Mode8 & r=0.008m, v=100.1m/s, Moving direction (x) \\
Mode9 & r=0.008m, v=100m/s, Moving direction (y) \\
Mode10 & r=0.008m, v=100.1m/s, Moving direction (y) \\
Mode11 & r=0.008m, v=100m/s, Moving direction (z) \\
Mode12 & r=0.008m, v=100.1m/s, Moving direction (z) \\
\hline
\end{tabular}
\end{table}

For the classification problem, Matconvnet \cite{Ref19}, a MATLAB toolbox was used. First, the crack patterns were stored as the dump files through the MD simulation based on the peridynamic theory. These dump files were converted to the 64 $\times$ 64 Numpy array and saved as the data. There are three types of data: training data set, validation data set, and testing data set. The total number of dump files is 10800, and 12 modes of data were obtained by changing the four parameters mentioned in section 4. The 12 modes are represented in table 1. We also randomly shuffled all data to reduce the imbalance of the data and also correctly divided the number of data according to each mode and assigned them to do training, validation, and test data set. Also, 30\% of the training data was designated as a validation data set to observe the overfitting and evaluate the training. Through these processes, we have used 6300 training data sets, 2700 validation data sets, and 1800 test data sets. 
In neural processes, PyTorch \cite{Ref20}, a deep learning implementation library, was used for the training and the testing, and the crack dataset was made the same as the MNIST \cite{Ref21} data format. Through the MD simulation using peridynamic theory, 10800 dump files were obtained as output. These are converted to the Numpy array and converted to 28 $\times$ 28 png image files. Finally, the training data set, the training label, the test data set, and the test label are stored separately and then converted into the ubyte.gz file to obtain the data and the label similar to MNIST dataset. Through these processes, 9000 training data, 9000 training labels, 1080 test data, and 1080 test labels are obtained.

\subsection{The composition of data sets for the CNNs and the Neural Processes}
\label{sec:5.2}

For the convolutional neural networks, the training data set and the test data set were made into a .mat file to apply the data for Matconvnet. The .mat file consists of `images' and `meta' format. First of all, the `images' part has a structure of `labels', `set', and `data'. The `labels are used to configure the labels for randomly mixed data. In `labels', our dataset is labeled randomly from 1 to 12 to classify as the 12 modes. `Set' specifies the training data set, test data set, and validation data set. For example, 1, 2, and 3 can be referred to as training data, test data, and validation data, respectively. The `data' stores the images of the crack patterns with Numpy array in the form of a 4-D single. For instance, if the size of 4-D single is 64 $\times$ 64 $\times$ 1 $\times$ 10800, there are 10800 grey images with 64 $\times$ 64 pixels. The `meta' structure contains `set' and `classes'. The `set' is related to `set' in `images', where 1 is training data, 2 is test data, and 3 is validation data, respectively. Finally, the `classes' are specified from 1 to 12 and related to the modes of the data we want to classify. 
Also, our dataset was made similar to the MNIST data type to apply the neural processes. All data were compressed in ubyte.gz in the form of training sets, training labels, test sets, and test labels, respectively. Likewise, the data were randomly mixed before storing the data and labels. Since the validation data set does not set up yet, it is possible to specify the validation data set to use a portion of the entire training set. The labels exist from 1 to 12, which are the same as the number of the classification. The data set was arranged to be easy to see at a glance by attaching the side of each image by side with 28 $\times$ 28 pixels to seem like the MNIST data type. All of these data types are ndarrays, an N-dimensional array.

\subsection{Convolutional neural networks for classification}
\label{sec:5.3}

\begin{figure}
\includegraphics[scale=0.33]{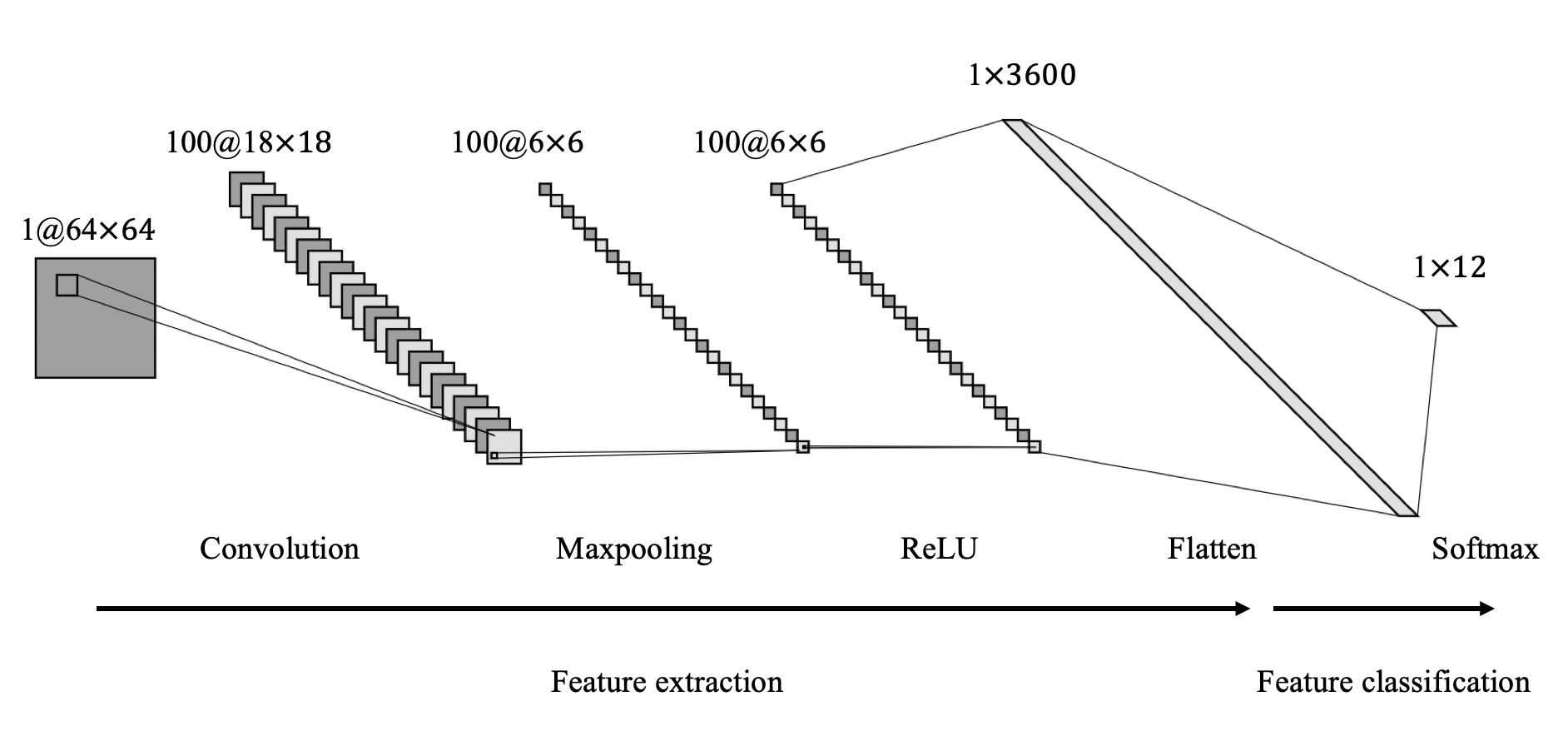}
\caption{Architecture of CNNs for classification of modes.}
\label{fig:4}       
\end{figure}

Deep learning has become a powerful tool for machine learning, and it is possible to perform various cognition and inference by appropriately utilizing the human knowledge existing in annotated data. CNNs are widely used to solve many computer vision problems such as image classification and object recognition etc. \cite{Ref22,Ref23,Ref24}. These neural networks are composed of several layers of neurons between the input and output, which can be shared with neurons in other layers, making it very easy to communicate information. Each connection unit acts as a linear or nonlinear operator defined by the set of parameters. We used the Alexnet \cite{Ref25} structure to classify the modes of the crack patterns. There are two processes for classifying properly annotated data. The first is the feature extraction step. This step takes a gray 2D image of a $\times$ a pixels as an input. This input image forms a 1 $\times$ c feature vector through each layer of neurons. The second is the feature classification. This process is the process of classifying the feature vector into the desired 12 modes. Through these two processes, images can be classified. Let's take a look at this process in more detail. First, the process of the feature extraction step requires three layers. The first is the convolutional layer. The convolutional layer receives the input image of $W_1$ $\times$ $H_1$ $\times$ $D_1$ and produces an output image of size $W_2$ $\times$ $H_2$ $\times$ $D_2$. Four hyperparameters are required in this process. Hyperparameters consist of the horizontal and vertical spatial size (F), stride (S), zero-padding (P), and the number of the filter (K). The input image forms an output image by the relation of four hyperparameters, and the relation equations are as follows.

\begin{equation}
W_2=(W_1−F+2P)/S+1
\end{equation}

\begin{equation}
H_2=(H_1−F+2P)/S+1
\end{equation}

\begin{equation}
D_2=K
\end{equation}

\noindent The volume of the output image is determined through these relations. Let's describe our calculations through Fig. 4. When an input image of 64 $\times$ 64 $\times$ 1 passes through the convolutional layer using the above-mentioned relation equations with hyperparameters of 11 $\times$ 11 filter size and 100 filters, stride (1) and zero padding (0), output images of the 18 $\times$ 18 $\times$ 100 are obtained. The second layer is the Max pooling layer \cite{Ref25}. This layer receives the input image passed through the convolutional layer and downsamples it. Therefore, it reduces the computational load and memory usage and makes calculations faster. We set the Max pooling to 3 in this study. Thus, applying the Max pooling layer to an image of 18 $\times$ 18 $\times$ 100 obtained through the convolutional layer, input images are converted to 6 $\times$ 6 $\times$ 100. Finally, it is the ReLU layer \cite{Ref27}, which is an activation function that is applied to the neuron at each pixel location. Also, by changing all values less than 0 to 0, the computation speed is significantly increased. The ReLU activation function is defined as follows:

\begin{equation}
ReLU(x) = max(0,x)
\end{equation}

\noindent where x is the intensity of a pixel. It controls how much information is passed through the network. So far, we have looked at the process of feature extraction. Finally, when we look at feature classification, the output image of 6 $\times$ 6 $\times$ 100 obtained through the three layers mentioned above was arranged in the form of 1 $\times$ 3600 vectors through the flattening process. The reason is that our ultimate goal is to classify into 12 modes. Thus, a flattened 1 $\times$ 3600 vectors is classified as 1 $\times$ 12 through the SoftMax layer \cite{Ref28}. SoftMax layer is often utilized in the final layer of a neural network-based classifier. It takes a vector of arbitrary real-valued scores (in z) and reduces it to a vector of values between 0 and 1. This process is as follows:

\begin{equation}
z^{[L]}=w^{[L]}a^{[L-1]}+b^{[L]}
\end{equation}

\noindent To compute z, we need SoftMax activation function

\begin{equation}
t = exp^{z^{[L]}}
\end{equation}

\noindent t, $z^{[L]}$ = (12,1) dimensional vector related to the classification for 12 modes.

\begin{equation}
a^{[L]}=\frac{exp^{z^{[L]}}}{\displaystyle\sum_{i=1}^{12}t_i}
\end{equation}
\noindent Output(a) is going to be the vector t but normalized to sum to 1.

\subsection{Results of training and testing using CNNs with success rates}
\label{sec:5.4}

\begin{figure}
\includegraphics[scale=0.45]{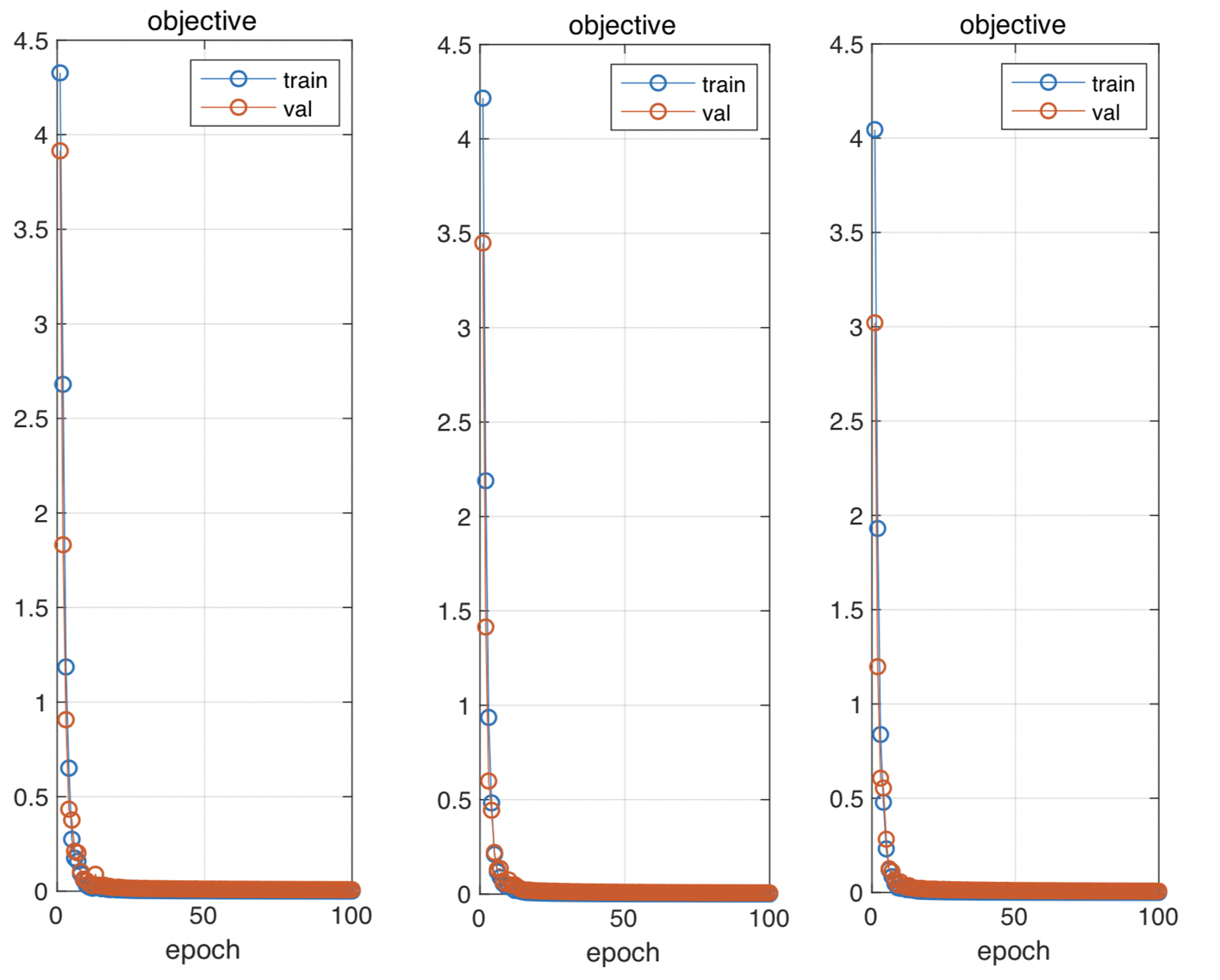}
\caption{The objective plots for PMB, LPS and VES model.}
\label{fig:5}       
\end{figure}

\begin{table}[ht]
\centering
\caption{CNNs prediction results compared to the test image data.}
\label{tab:2}       
\begin{tabular}{c|c}
\hline
\hline
The PMB test image number : Label(Mode) & CNNs prediction results (Mode)  \\
\hline
Test image 1 : 4 (Mode) & 4 (Mode) \\
Test image 2 : 11 (Mode) & 11 (Mode) \\
Test image 3 : 5 (Mode) & 5 (Mode) \\
Test image 4 : 3 (Mode) & 3 (Mode) \\
Test image 5 : 7 (Mode) & 7 (Mode) \\
Test image 6 : 2 (Mode) & 2 (Mode) \\
Test image 7 : 4 (Mode) & 4 (Mode) \\
Test image 8 : 8 (Mode) & 8 (Mode) \\
Test image 9 : 1 (Mode) & 1 (Mode) \\
Test image 10 : 2 (Mode) & 2 (Mode) \\
\hline
The LPS test image number : Label(Mode) & CNNs prediction results (Mode)  \\
\hline
Test image 1 : 10 (Mode) & 10 (Mode) \\
Test image 2 : 7 (Mode) & 7 (Mode) \\
Test image 3 : 3 (Mode) & 3 (Mode) \\
Test image 4 : 5 (Mode) & 5 (Mode) \\
Test image 5 : 2 (Mode) & 2 (Mode) \\
Test image 6 : 6 (Mode) & 6 (Mode) \\
Test image 7 : 1 (Mode) & 1 (Mode) \\
Test image 8 : 12 (Mode) & 12 (Mode) \\
Test image 9 : 4 (Mode) & 4 (Mode) \\
Test image 10 : 5 (Mode) & 5 (Mode) \\
\hline
The VES test image number : Label(Mode) & CNNs prediction results (Mode)  \\
\hline
Test image 1 : 5 (Mode) & 5 (Mode) \\
Test image 2 : 8 (Mode) & 8 (Mode) \\
Test image 3 : 3 (Mode) & 3 (Mode) \\
Test image 4 : 10 (Mode) & 10 (Mode) \\
Test image 5 : 9 (Mode) & 9 (Mode) \\
Test image 6 : 7 (Mode) & 7 (Mode) \\
Test image 7 : 5 (Mode) & 5 (Mode) \\
Test image 8 : 3 (Mode) & 3 (Mode) \\
Test image 9 : 2 (Mode) & 2 (Mode) \\
Test image 10 : 11 (Mode) & 11 (Mode) \\
\hline
\end{tabular}
\end{table}

In this section, we examined the results of training, validation, testing, and success rates in classifying modes using specified data and CNNs. As mentioned earlier, in the 9000-training dataset, 30$\%$ of training data were used for the validation dataset in the training process. Fig. 5 shows the training and validation results according to the PMB, LPS, and VES models. During the training, hyperparameters were set such as the learning rate was 0.001, the batch size was 50, and the number of the epoch was 100. As we can see in Fig. 5, we observed the objective value and judged whether the training was proper or not. The blue line represents a training error, and the orange line indicates a validation error.
The objective value is the same as the error graph, and there is some error when the training is first started. However, as the epoch increases, it gradually converges to zero. It is also possible to judge the overfitting of the training through Fig. 5. If there is overfitting in the training process, the validation line may not converge to 0, or the value may drop, and it may converge to a particular value. In this case, we can make the model simpler, or reduce the overfitting by regularization. In this study, as shown in Fig. 5, training and validation errors converge to zero precisely without overfitting or underfitting. So far, we have studied the training and validation process. Finally, we tested the 1200 images based on the training data and judged how accurately we predicted the modes and classified the test image through the success rates.

\noindent Table. 2 represents the classification results according to the models using the training data. The test image number and Label (Mode) indicate the image data we tested, and the labels associated with it. CNNs prediction results (Mode), on the other hand, show what the test images predict.
In the results, the prediction of the test results according to the model was successful. We obtained the success rate, and the relation equation is as follows:

\begin{equation}
\frac{number \, of \, the\, correct\, test\, images}{the\, total\, number\, of\, the\, test\, images} \times 100\%
\end{equation}

\noindent The prediction results of the classification modes using the training data showed the success rates between 99.6$\%$ and 99.8$\%$.

\subsection{The 2-D regression problem results using Neural Processes }
\label{sec:5.5}

\begin{figure}
  \centering
  \begin{tabular}{c}
  \includegraphics[scale=0.143]{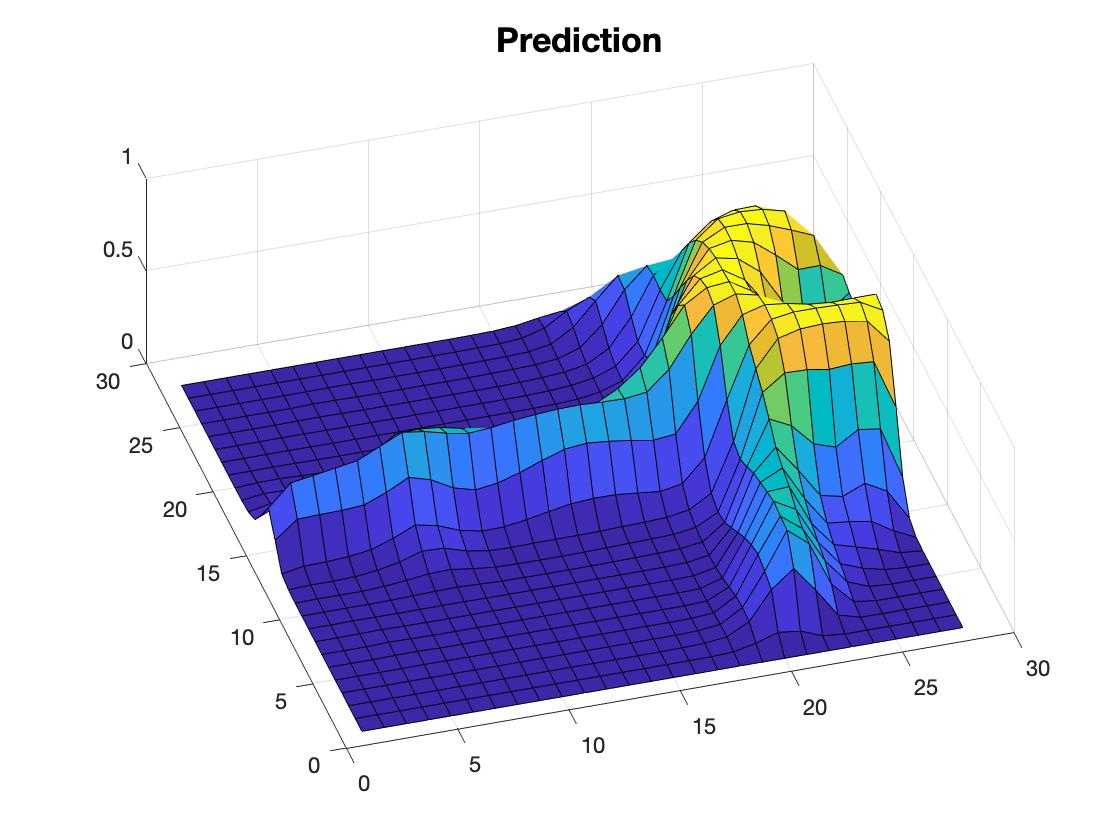} 
  \includegraphics[scale=0.143]{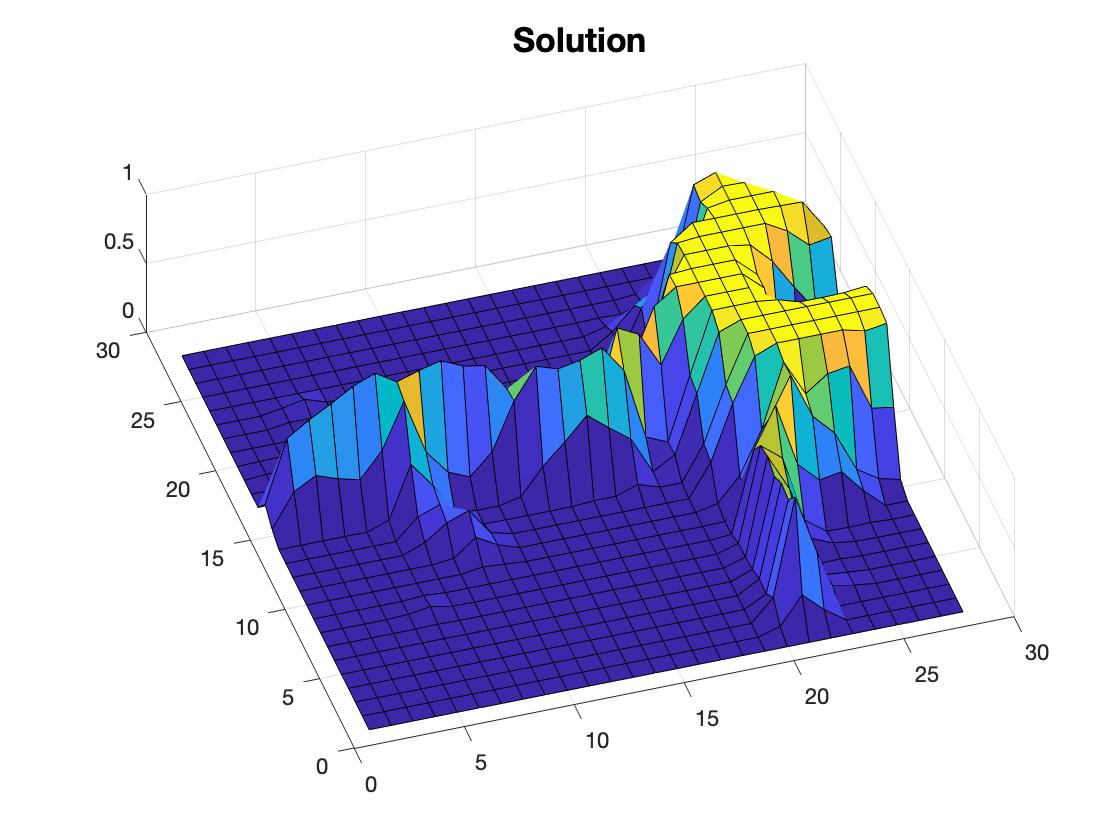} \\ 
  \includegraphics[scale=0.115]{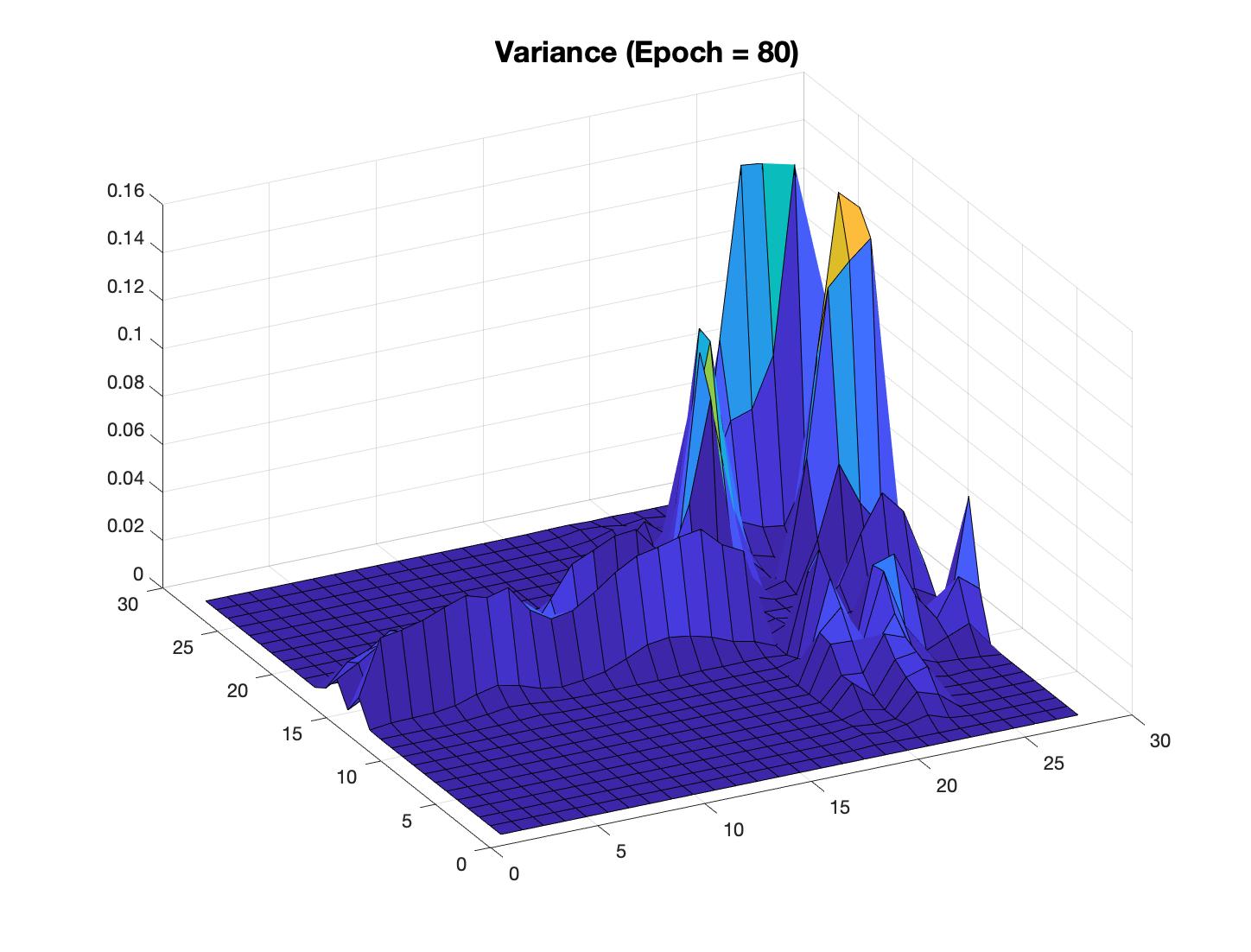}
  \includegraphics[scale=0.115]{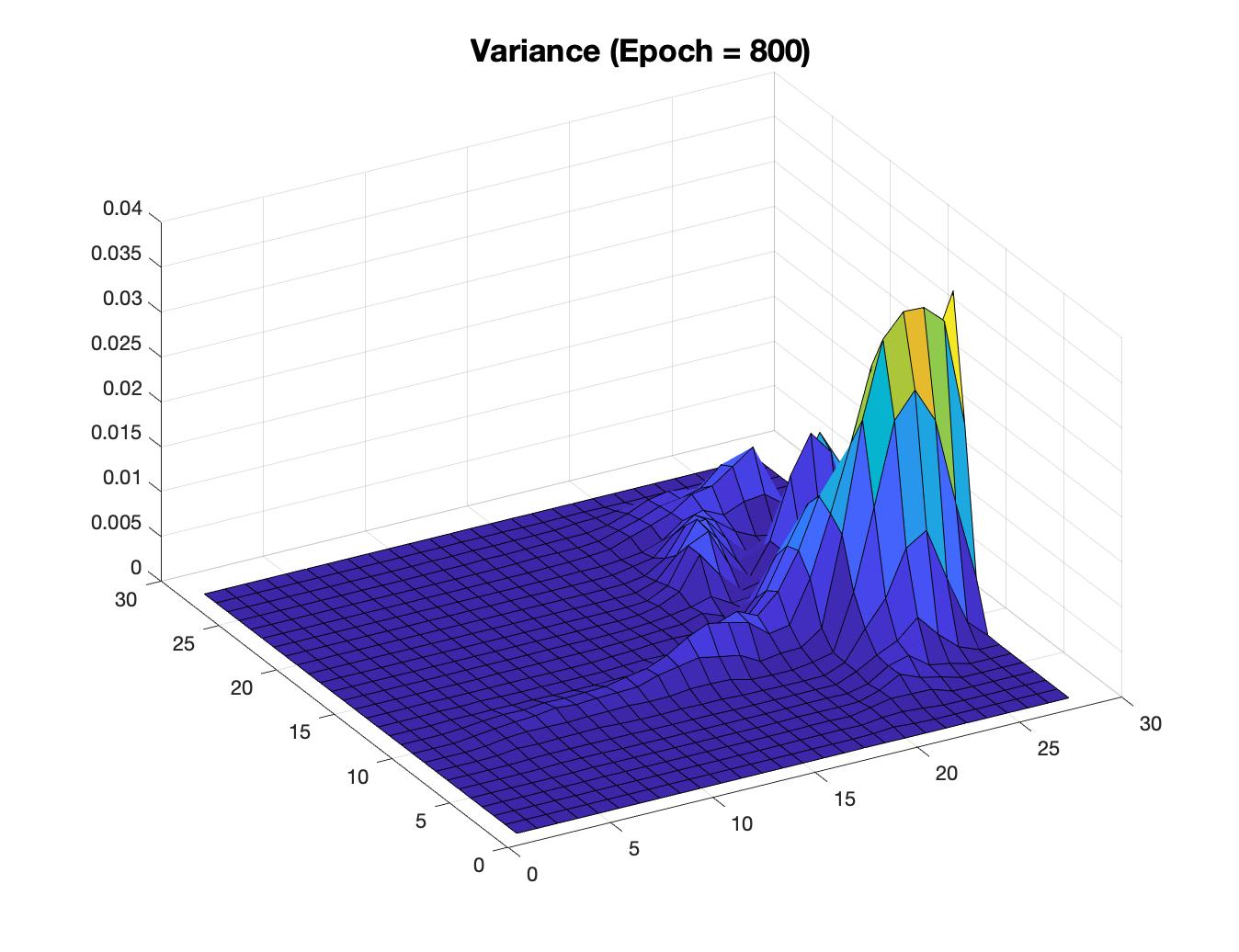}
  \end{tabular}
\caption{Comparison between the NPs' prediction and the true solution for the crack pattern problem. The prediction and solution on the top have been obtained from the test data set and have not been considered by the NPs during the training procedure. The prediction and the true solution are represented on the top, and the bottom figures show the variance at different epoch numbers.}  
\label{fig:6}       
\end{figure}

\begin{figure}
  \centering
  \begin{tabular}{c}
  \includegraphics[scale=0.4]{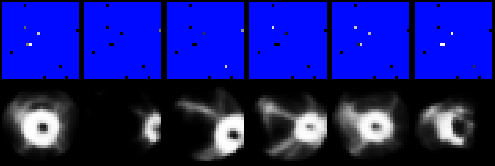} \\
  \includegraphics[scale=0.4]{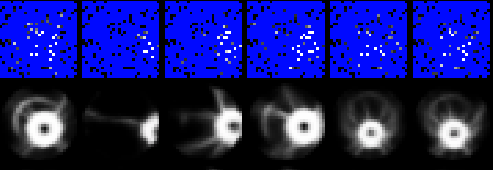} \\
  \includegraphics[scale=0.4]{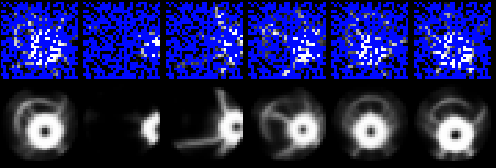} \\
  \includegraphics[scale=0.4]{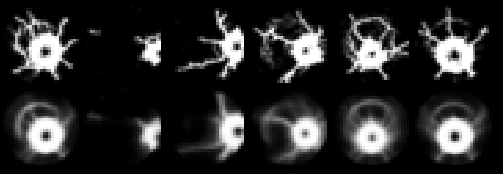}
  \end{tabular}
\caption{2-D regression problem results for crack patterns with 10, 100, 300, 784 contexts points at epoch number = $3,000$. 1) The first image represents the prediction results with 10 context points. 2) The second image represents the prediction results with 100 context points. 3) The third image represents the prediction results with 300 context points. 4) The last image represents the prediction results with 784 context points.
 }
\label{fig:7}       
\end{figure}

So far, we have classified the crack patterns according to the modes by CNNs. In this section, we try to predict the crack patterns by solving the 2-D regression problem using NPs. As mentioned earlier, the data used 9000 training data and 1800 test data. The structure of the data is similar to that of MNIST, and the size of each image is 28 $\times$ 28 pixels. The NPs were used to solve the 2-D regression problem, and the experiment set up of the NPs was as follows. In hyperparameters, the input batch size for training was set to 128, and the number of the epoch was set to 3000. Also, the dimension of representation (r) was 128 to obtain each representation through the NNs from context points. This is associated with the linear layers of NNs which yield the 128 representations through three linear layers (3 $\times$ 400, 400 $\times$ 400 and 400 $\times$ 128). In NNs, we used the ReLU layer as an activation function and added efficiency to the training. Representation obtained through NNs is used to parametrize the distribution of z, which is related to the Eq. (17), and through this process, the dimension of the global latent variable (z) is also 128.

Using the obtained z and target points, we predicted the crack patterns through the g of the linear layers. The linear layers of the g are composed of 130 $\times$ 400, 400 $\times$ 400, 400 $\times$ 400, 400 $\times$ 1. As a result, crack patterns can be predicted.
Finally, the loss function is binary cross-entropy, and the relation equation is as follows:

\begin{equation}
H_P(q) = -\frac{1}{N}{\displaystyle\sum_{i=1}^{N}y_i\cdot \log(p) + (1-y_i)\cdot 
\log(1-p)}
\end{equation}

\noindent For instance, considering the modes, we assume that there are modes 1 and 2. y represents the label for mode 1, and p is the predicted probability of the point being mode 1 for all N points. This represents that for each mode 1 point, it adds $\log$(p) to the loss, that is, the log probability of it being mode 1. On the contrary, it adds $\log$(1-p), that is, the log probability of it being mode 2, for each mode 1 point. Finally, the model was trained using standard ADAM optimization algorithms and backpropagation implemented in PyTorch. The learning speed of the ADAM optimization was set to 0.001. 

In this study, the network predictions were evaluated using the above network settings, and the results are shown in Fig. 6. The network prediction and true solution can be found at the top of the line and can be seen to be quite similar when comparing the results. In the next line, 
we represented the results of the variance obtained from the neural processes as increasing the epoch numbers.
As a result of the variance in Fig. 6, the variance value was significantly lowered when the process of the prediction proceeded from epoch 1 to reach the epoch 800, and the range was from 0 to 0.035.
We can see from the results that the prediction value becomes similar to the true solution value because the variance value becomes smaller according to the result as increasing the epoch iterations and parameter tuning.
The variance can be used as an indicator to show how good the network’s prediction is. Also, the location with the most massive variance is the location where we should take additional samples. We can use it for experimental design to get additional optimal sample locations.
Using the structure of the generated neural processes, the predicted results obtained from one of our models, the LPS model, through the training and testing, are shown in Fig. 7. The figure shows the results of the prediction of the test data. In the training process, we set the context points to 10, 100, 300, and 784 and represent the test results accordingly in 3000 epochs. The first figure shows the result of the 3000 epochs according to the context points 10. As you can see from the first figure, the test images are not predicted well since the information of the test images is not enough to predict the data in the training process because of setting the 10 context points. 
The second figure shows the results of anticipating the test images with the 100 context points. In this process, the crack pattern is blurred, but the crack pattern can be predicted gradually. Finally, the last image illustrates the results of anticipating the test images after finishing the final training at epoch 3000 with 784 context points. As we can be seen from this result, we accurately predicted the crack patterns and even more surprising is that also if the context points are 100 and 300 the crack patterns are predicted precisely as well as the results with the context points of 784. Through this process, if NPs are used, even if there is not enough information, it is enough to predict the model.

\section{Conclusion}
\label{sec:6}

We used the peridynamics to obtain the crack patterns changing the four variables: the velocity of the indenter, the radius of the indenter, the hitting location of the disk, and the moving direction of the disk. Peridynamics can compensate for the disadvantages of FEM and generate more accurate crack patterns.
In addition, various models (PMB, LPS, VES) could be complemented by using case studies with convolutional neural networks using peridynamics. First, the modes of the crack pattern were classified, and the case study of models was conducted with convolutional neural networks. In this process, data obtained through the peridynamics have labeled the data according to each mode and did the supervised learning. By evenly distributing the data, the problem of the imbalanced data was reduced. A training data set of 30$\%$ was designated as a validation data set to solve the most problematic overfitting in the training process. Overfitting was observed during the training and training was optimized by tuning hyperparameters to reduce the overfitting. As a result, 99.6$\%$ $\sim$ 99.8$\%$ success rate was obtained, and the modes could be classified accurately.

Finally, neural processes were used to solve the 2-D regression problem. These neural processes were optimized to address the regression problem by combining the advantages of neural networks and Gaussian processes. In this process, the data obtained through the peridynamics were made similar to the MNIST data and were classified into training data set and test data set for training. We also set the context points to 10, 100, 300, and 784 so that we could see the results of the test images. As a result, if the epoch is low, the prediction is not good enough regardless of context points. However, as the number of epochs increases, the result of the speculation is right. In the final epoch 3000, even for the case of $100$ context points, the prediction is sufficiently good as in the case of $784$ context points. In other words, using neural processes, the forecast is possible using data with less information.

\begin{acknowledgements}
We gratefully acknowledge the support from the National Science Foundation (DMS-1555072, and DMS-1736364).
\end{acknowledgements}


\begin{thebibliography}{}

\bibitem{Ref1}
Seleson, P., Parks, M. L., Gunzburger, M., \& Lehoucq, R. B. (2009). Peridynamics as an upscaling of molecular dynamics. Multiscale Modeling \& Simulation, 8(1), 204-227.

\bibitem{Ref2}
Silling, S., Epton, A., Weckner, M., Xu, O., \& Askari, J. (2007). Peridynamic States and Constitutive Modeling. Journal of Elasticity, 88(2), 151-184.

\bibitem{Ref3}
Silling, S. (2000). Reformulation of elasticity theory for discontinuities and long-range forces. Journal of the Mechanics and Physics of Solids, 48(1), 175-209.

\bibitem{Ref4}
Bobaru, F., Silling, S. A., \& Jiang, H. (2005). Peridynamic fracture and damage modeling of membranes and nanofiber networks. In XI Int. Conf. Fract., Turin, Italy.

\bibitem{Ref5}
Askari, E., Xu, J., \& Silling, S. (2006). Peridynamic analysis of damage and failure in composites. In the 44th AIAA aerospace sciences meeting and exhibit (p. 88).

\bibitem{Ref6}
Nikabdullah, N., Azizi, Alebrahim, Singh, and K. "The Application of Peridynamic Method on Prediction of Viscoelastic Materials Behaviour." AIP Conference Proceedings 1602.1 (2014): 357-63. Web.

\bibitem{Ref7}
Plimpton, S. (1995). Fast Parallel Algorithms for Short-Range Molecular Dynamics. Journal of Computational Physics, 117(1), 1-19.

\bibitem{Ref8}
M.L. Parks, D.J. Littlewood, J.A. Mitchell, and S.A. Silling, Peridigm Users’ Guide, Tech. Report SAND2012-7800, Sandia National Laboratories, 2012.

\bibitem{Ref9}
Silling, S., \& Askari, E. (2005). A meshfree method based on the peridynamic model of solid mechanics. Proposed for publication in Computers and Structures., 83(17-18), Proposed for publication in Computers and Structures., 2005, Vol.83(17-18).

\bibitem{Ref10}
Parks, M. L., Seleson, P., Plimpton, S. J., Silling, S. A., \& Lehoucq, R. B. (2011). Peridynamics with lammps: A user guide, v0. 3 beta. Sandia Report (2011–8253), 3532.

\bibitem{Ref11}
Mitchell. A non-local, ordinary-state-based viscoelasticity model for peridynamics. Sandia National Lab Report, 8064:1-28 (2011).

\bibitem{Ref12}
Kim, M., Winovich, N., Lin, G., \& Jeong, W. (2019). Peri-Net: Analysis of Crack Patterns Using Deep Neural Networks. Journal of Peridynamics and Nonlocal Modeling, 1(2), 131-142.

\bibitem{Ref13}
Garnelo, M., Schwarz, J., Rosenbaum, D., Viola, F., Rezende, D. J., Eslami, S. M., \& Teh, Y. W. (2018). Neural processes. arXiv preprint arXiv:1807.01622.

\bibitem{Ref14}
Parks, M. L., Lehoucq, R. B., Plimpton, S. J., \& Silling, S. A. (2008). Implementing peridynamics within a molecular dynamics code. Computer Physics Communications, 179(11), 777-783.

\bibitem{Ref15}
Rasmussen, C. E. (2004). Gaussian processes in machine learning. Advanced lectures on machine learning (pp. 63-71). Springer, Berlin, Heidelberg.

\bibitem{Ref16}
S. J. Plimpton, Pizza.py http://www.cs.sandia.gov/ sjplimp/pizza.html.

\bibitem{Ref17}
Ahrens, James, Geveci, Berk, Law, Charles, ParaView: An End-User Tool for Large Data Visualization, Visualization Handbook, Elsevier, 2005, ISBN-13: 978-0123875822

\bibitem{Ref18}
Ayachit, Utkarsh, The ParaView Guide: A Parallel Visualization Application, Kitware, 2015, ISBN 978-1930934306
\bibitem{Ref19}
Vedaldi, Andrea, and Karel Lenc. "MatConvNet - Convolutional Neural Networks for MATLAB." (2014). Web.

\bibitem{Ref20}
Paszke, A., Gross, S., Chintala, S., Chanan, G., Yang, E., DeVito, Z., ... \& Lerer, A. (2017). Automatic differentiation in PyTorch.

\bibitem{Ref21}
LeCun, Y., Bottou, L., Bengio, Y., \& Haffner, P. (1998). Gradient-based learning applied to document recognition. Proceedings of the IEEE, 86(11), 2278-2324.

\bibitem{Ref22}
Goodfellow, I., Bengio, Y., Courville, A., \& Bengio, Y. (2016). Deep learning (Vol. 1). Cambridge: MIT Press.

\bibitem{Ref23}
Géron, A. (2017). Hands-on machine learning with Scikit-Learn and TensorFlow: concepts, tools, and techniques to build intelligent systems. " O'Reilly Media, Inc.".
\bibitem{Ref24}
LeCun, Y., Bengio, Y., \& Hinton, G. (2015). Deep learning. nature, 521(7553), 436.


\bibitem{Ref25}
Krizhevsky, A., Sutskever, I., \& Hinton, G. E. (2012). Imagenet classification with deep convolutional neural networks. In Advances in neural information processing systems (pp. 1097-1105).

\bibitem{Ref26}
Riesenhuber, M., \& Poggio, T. (1999). Hierarchical models of object recognition in cortex. Nature Neuroscience, 2(11), 1019.

\bibitem{Ref27}
Krizhevsky, A., Sutskever, I., \& Hinton, G. E. (2012). Imagenet classification with deep convolutional neural networks. In Advances in neural information processing systems (pp. 1097-1105).

\bibitem{Ref28}
Nasrabadi, N. M. (2007). Pattern recognition and machine learning. Journal of electronic imaging, 16(4), 049901.

\end{thebibliography}
\end{document}